# Echo State Learning for Wireless Virtual Reality Resource Allocation in UAV-enabled LTE-U Networks

Mingzhe Chen*, Walid Saad†, and Changchuan Yin*

*Beijing Laboratory of Advanced Information Network,
Beijing University of Posts and Telecommunications, Beijing, China 100876, Emails: chenmingzhe@bupt.edu.cn, ccyin@ieee.org.
†Wireless@VT, Bradley Department of Electrical and Computer Engineering, Virginia Tech, Blacksburg, VA, USA, Email: walids@vt.edu.

*Abstract*—In this paper, the problem of resource management is studied for a network of wireless virtual reality (VR) users communicating using an unmanned aerial vehicle (UAV)-enabled LTE-U network. In the studied model, the UAVs act as VR control centers that collect tracking information from the VR users over the wireless uplink and, then, send the constructed VR images to the VR users over an LTE-U downlink. Therefore, resource allocation in such a UAV-enabled LTE-U network must jointly consider the uplink and downlink links over both licensed and unlicensed bands. In such a VR setting, the UAVs can dynamically adjust the image quality and format of each VR image to change the data size of each VR image, then meet the delay requirement. Therefore, resource allocation must also take into account the image quality and format. This VR-centric resource allocation problem is formulated as a noncooperative game that enables a joint allocation of licensed and unlicensed spectrum bands, as well as a dynamic adaptation of VR image quality and format. To solve this game, a learning algorithm based on the machine learning tools of echo state networks (ESNs) with leaky integrator neurons is proposed. Unlike conventional ESN based learning algorithms that are suitable for discrete-time systems, the proposed algorithm can dynamically adjust the update speed of the ESN's state and, hence, it can enable the UAVs to learn the continuous dynamics of their associated VR users. Simulation results show that the proposed algorithm achieves up to 14% and 27.1% gains in terms of total VR QoE for all users compared to Q-learning using LTE-U and Q-learning using LTE.

## I. INTRODUCTION

Virtual reality (VR) services are viewed as one of the primary use cases of tomorrow's 5G cellular networks [1]. Indeed, wireless VR services will enable users to experience immersive services, without the constraints imposed by conventional wired VR systems. To enable the users to experience immersive VR applications without any constraints, VR systems can be operated using wireless networking technologies [1]. However, deploying VR services over wireless cellular systems, such as LTE and beyond, faces many challenges [1] that range from providing reliable data transmission to optimizing tracking and resource allocation.

Some recent works such as in [1]–[5] have studied a number of problems related to wireless VR. In [1] and [2], the authors provided a qualitative summary of the challenges of VR over wireless. However, these works rely on simple toy examples, and do not provide a rigorous analytical treatment of the wireless networking challenges of VR. The authors in [3] proposed a channel access scheme using the unlicensed band for wireless multi-user VR system. The work in [4] investigated the use of VR devices to control an unmanned drone with 2.4 GHz Wi-Fi grid. However, the works in [3] and [4] only focus on how to use WiFi technology to transmit VR images and do not consider the resource allocation problem over downlink and uplink, licensed and unlicensed bands for VR over cellular networks. In [5], we proposed a wireless VR model that captures the tracking accuracy, processing delay, and transmission delay and introduced a machine learning algorithm for resource allocation. However, our work in [5] is restricted to classical LTE networks in which the VR devices and static ground base stations can access only a single, licensed band. However, relying only on the licensed band may not allow the cellular base stations to meet the high data rate demands of the VR users. One potential solution is to exploit the use of LTE over unlicensed (LTE-U) [6] bands Moreover, in dense urban environments, the presence of high rise buildings and obstacles, can present a major challenge for meeting the delay needs for VR transmissions by using conventional, terrestrial base stations. Instead, a flying base station carried by an unmanned aerial vehicle (UAV) [7] that can move to an optimal location to establish communication links with little blockage can be considered to provide better connectivity and assist any ground network. Moreover, all of the existing works such as in [1]–[5] ignore the image quality and image format [8] in the improvement of the delay of the VR users. Indeed, as the VR image quality and format change, the data size of each image will be varied and, hence, wireless base stations will need to dynamically adjust the image' quality and format to effectively service the users.

The main contribution of this paper is to introduce a novel framework for enabling VR applications over UAV-based LTE-U networks. To the best of our knowledge, *this is the first work that jointly considers licensed and unlicensed spectrum resource allocation, image format and quality allocation for VR users over UAV-based LTE-U networks.* Hence, our key contributions include:

- For the considered VR applications over a wireless LTE-U network, we develop an effective resource allocation scheme that allocates licensed and unlicensed resource blocks over uplink and downlink and adjusts the image quality and format for each VR user's image. The goal of this scheme is to maximize the quality-of-experience (QoE) of all users while meeting the delay requirement of the VR system. We formulate the problem as a noncooperative game in which the players are the UAVs. Each player seeks to find an optimal resource allocation scheme to optimize the users' QoE while meeting the delay requirement.
- To solve this game, we propose a learning algorithm based on echo state networks (ESNs) with leaky integrator neurons [9]. This ESN-based algorithm can effectively find a Nash equilibrium of the game. The proposed algorithm enables the UAVs to learn the continuous dynamics of their associated VR users hence allowing adaptation to environmental dynamics.
- Simulation results show that the proposed algorithm can, respectively, achieve 14% and 27.1% gains in terms of total

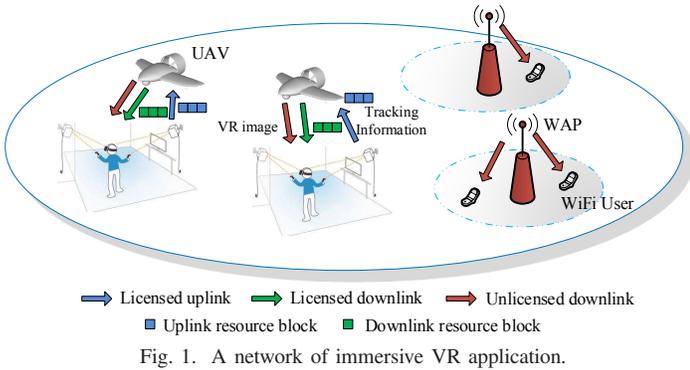

Fig. 1. A network of immersive VR application.

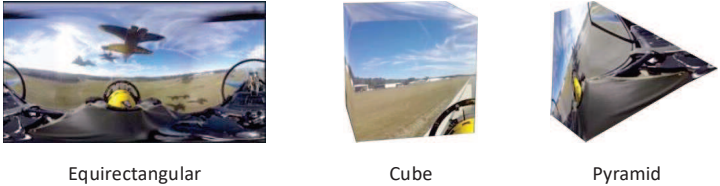

Equirectangular         Cube         Pyramid

Fig. 2. The formats of a VR image.

VR QoE for all users compared to Q-learning using LTE-U network and Q-learning using LTE network.

The rest of this paper is organized as follows. The system model and problem formulation are presented in Section II. The ESN-based resource allocation algorithm is proposed in Section III. In Section IV, numerical simulation results are presented and analyzed. Finally, conclusions are drawn in Section V.

## II. SYSTEM MODEL AND PROBLEM FORMULATION

Consider the downlink and uplink transmission of an LTE-U network composed of a set $\mathcal{B}$ of $B$ unmanned aerial vehicles (UAVs) and $W$ ground WiFi access points (WAPs). The UAVs are deployed to act as flying LTE-U base stations to serve a set $\mathcal{U}$ of $U$ wireless VR users. Here, we consider *dual-mode* UAVs that are able to access both the licensed and unlicensed bands. The licensed band is used for both downlink and uplink transmission while the unlicensed band is only used for the downlink transmission (due to the more stringent downlink traffic). The downlink is used to transmit the *VR images* displayed on each user's VR device while the uplink is used to transmit the *tracking information* that is used to construct a user's VR image as shown in Fig. 1. Here, the tracking information includes a user's orientation and location information. In our model, the UAVs adopt an orthogonal frequency division multiple access (OFDMA) technique over the licensed band and transmit over uplink resource blocks and downlink resource blocks. Our LTE-U model uses a time division duplexing (TDD) mode with duty cycle method [6]. Under this method, LTE-U and WiFi users will, separately, use the time slots on the unlicensed band. In particular, UAVs will use a fraction $\vartheta$ of time for VR transmission and will be muted for $1-\vartheta$ time which is allocated to the WiFi transmission links.

The coverage of each UAV is considered to be a circular area with radius $r$ and each UAV only allocates resource blocks to the users located in its coverage range. Each user can only connect to a single UAV. Note that, in our model, we consider static users that move only within confined areas, as is the case for typical VR entertainment applications, such as immersive videos.

### A. VR Image Transmission

The data size of each VR image depends on the quality and the format of each VR image. The quality of a VR image can be captured by the resolution of a VR image (720p, 1080p, etc.). When the resolution of each VR image increases, the number of pixels that each UAV needs to construct the VR image will also increase. We use a set $\mathcal{L} = \{l_1, l_2, \ldots, l_K\}$ to represent the level of VR image quality where $l_i \in \mathcal{L}$ indicates that the quality of a VR image is at level $i$ and $l_K$ represents the highest quality of a VR image [10] ($l_K > l_i, i = 1, 2, \ldots, K-1$). In VR, the standard format of a VR image is equirectangular [8]. However, to reduce its data size, a VR image can be transformed from the standard equirectangular format into either cubic or pyramidal formats as shown in Fig. 2. Using the cube or pyramid formats can reduce, respectively, 25% and 80% of data size compared to using the equirectangular format [8]. However, reducing the data size of a VR image will decrease the QoE of each user. This is due to the fact that as the data size of a VR image decreases, the number of the pixels used to construct a VR image decreases [8] and, hence, the users may not be able to see the entire 360° VR image. We use a set $\mathcal{F} = \{f_1, f_2, f_3\}$ to represent the formats of a VR image. Here, $f_1$ to $f_3$, respectively, represent the pyramid, cube, and standard formats. Based on the above formulation, for user $i$, the data size of a VR image is defined as $m_i(l, f), l \in \mathcal{L}, f \in \mathcal{F}$. Note that $m_i(l_K, f_3)$ is the maximum data size of a VR image and $m(l_1, f_1)$ is the minimum data size of a VR image. The QoE of user $i$ can be given by $Q_i(l, f) = \frac{m(l,f)}{m(l_K, f_3)}$. Here, when the quality of a VR image increases, the QoE of a user will increase, which corresponds to the fact that watching a high definition video is more comfortable than watching a standard definition video.

### B. WiFi data rate analysis

For the considered WiFi system, a standard carrier sense multiple access with collision avoidance (CSMA/CA) protocol will be used along with its corresponding RTS/CTS access mechanism. The WAPs will use a CSMA/CA scheme with binary slotted exponential backoff. We assume that one LTE time slot consists of $T_W$ WiFi time slots. Based on the duty cycle mechanism, the UAVs can occupy a fraction $\vartheta$ of $T_W$ time slots over the unlicensed band while the WiFi users can occupy a fraction $(1-\vartheta)$ of $T_W$ time slots. Thus, the per WiFi user rate will be:

$$R_w = \frac{R(N)(1-\vartheta)}{N}, \quad (1)$$

where $R(N)$ is the saturation capacity of $N$ WiFi users sharing the same unlicensed band which is given by [6]. Given the rate requirement of each WiFi user $\xi$, the fraction of the unlicensed band time slot allocated to the LTE-U users can be given by $\vartheta \leq 1 - N\xi/R(N)$.

### C. Delay Model

In the studied system, VR images are transmitted from the UAVs to the users while tracking information is transmitted from the users to the UAVs. Hence, the transmission delay over the uplink and downlink will directly affect the users' QoE and the UAVs must guarantee the transmission delay for each user. In consequence, we must consider the transmission delay in this model. Next, we first introduce a probabilistic UAV channel model and, then, we formulate the transmission delay of each

VR user. In the considered UAV channel model, probabilistic line-of-sight (LoS) and non-line-of-sight (NLoS) links are considered as in [11]. NLoS links experience higher attenuations than LoS links due to the shadowing and diffraction loss [11]. The LoS and NLoS path loss of UAV $k$ transmitting a VR image to user $i$ over the licensed resource blocks can be given by (in dB) [12]:

$$g_{ij}^{\text{LoS}} = 20\log\left(\frac{4\pi d_{ij}f}{c}\right) + \eta_{\text{LoS}}^{\text{L}}, g_{ij}^{\text{NLoS}} = 20\log\left(\frac{4\pi d_{ij}f}{c}\right) + \eta_{\text{NLoS}}^{\text{L}},$$

where $20\log(d_{ij}f4\pi/c)$ is the free space path loss with $d_{ij}$ being the distance between user $i$ and UAV $j$, $f$ being the carrier frequency, and $c$ being the speed of light. Here, $\eta_{\text{LoS}}^{\text{L}}$ and $\eta_{\text{NLoS}}^{\text{L}}$ represent, respectively, additional attenuation factors due to the LoS/NLoS connections over the licensed resource blocks. The LoS probability is given by [12]:

$$\Pr\left(g_{ij}^{\text{LoS}}\right) = (1 + X\exp(-Y[\phi_{ij} - X]))^{-1}, \quad (2)$$

where $X$ and $Y$ are constants that depend on the environment and $\phi_{ij} = \sin^{-1}(h_j/d_{ij})$ is the elevation angle. Then, the average path loss from UAV $j$ to user $i$ will be [12]:

$$\bar{g}_{ij} = \Pr\left(g_{ij}^{\text{LoS}}\right) \times g_{ij}^{\text{LoS}} + \Pr\left(g_{ij}^{\text{NLoS}}\right) \times g_{ij}^{\text{NLoS}}, \quad (3)$$

where $\Pr\left(g_{ij}^{\text{NLoS}}\right) = 1 - \Pr\left(g_{ij}^{\text{LoS}}\right)$. The downlink rate of user $i$ associated with UAV $j$ on the licensed resource blocks is:

$$c_{ij}^{\text{L}}(\boldsymbol{q}_{ij}) = \sum_{k=1}^{Q} q_{ij,k} B\log_2(1 + \gamma_{ij,k}), \quad (4)$$

where $\boldsymbol{q}_{ij} = [q_{ij,1}, \ldots, q_{ij,Q}]$ is the vector of resource blocks that UAV $j$ allocates to user $i$ with $q_{ij,k} \in \{0,1\}$. Here, $q_{ij,k} = 1$ indicates that resource block $k$ is allocated to user $i$. $\gamma_{ij,k} = \frac{P_B \bar{g}_{ij}^k}{\sigma^2 + \sum_{l \in \mathcal{R}^k, l \neq j} P_B \bar{g}_{il}^k}$ is the downlink signal-to-interference-plus-noise ratio (SINR) between UAV $j$ and user $i$ over licensed resource block $k$. $\mathcal{R}^k$ represents the set of UAVs that use downlink resource block $k$, $B$ is the bandwidth of each resource block, $P_B$ is the transmit power of UAV $j$ which is assumed to be equal for all UAVs, and $\sigma^2$ is the variance of the Gaussian noise. For UAV $k$ transmitting a VR image to user $i$ using the unlicensed band, the LoS and NLoS path loss will be (in dB) [12]:

$$g_{ij}^{\text{LoS}} = 20\log\left(\frac{4\pi d_{ij}f}{c}\right) + \eta_{\text{LoS}}^{\text{U}}, g_{ij}^{\text{NLoS}} = 20\log\left(\frac{4\pi d_{ij}f}{c}\right) + \eta_{\text{NLoS}}^{\text{U}},$$

where $\eta_{\text{LoS}}^{\text{U}}$ and $\eta_{\text{NLoS}}^{\text{U}}$ represent additional attenuation factors due to, respectively, the LoS and NLoS connections over the unlicensed band. The LoS probability and the average path loss can be, respectively, calculated using a method that is similar to (2) and (3). Therefore, the downlink rate of user $i$ associated with UAV $j$ over the unlicensed band is:

$$c_{ij}^{\text{U}}(e_{ij}) = e_{ij}\vartheta B_u\log_2\left(1 + \gamma_{ij,k}^{\text{U}}\right), \quad (5)$$

where $B_u$ is the bandwidth of the unlicensed band, $e_{ij}$ is the fraction of $\vartheta$ over the unlicensed band with $\sum_i e_{ij} = 1$, and $\gamma_{ij,k}^{\text{U}}$ is the SINR between UAV $j$ and user $i$ over the unlicensed band. The total downlink rate of user $i$ associated with UAV $j$ is $c_{ij}(\boldsymbol{q}_{ij}, e_{ij}) = c_{ij}^{\text{L}}(\boldsymbol{q}_{ij}) + c_{ij}^{\text{U}}(e_{ij})$. The uplink rate of user $i$ associated with UAV $j$ can be given by:

$$c_{ij}(\boldsymbol{v}_{ij}) = \sum_{k=1}^{V} v_{ij,k} B\log_2\left(1 + \gamma_{ij,k}^{\text{u}}\right), \quad (6)$$

where $\boldsymbol{v}_{ij} = [v_{ij,1}, \ldots, v_{ij,V}]$ is the vector of resource blocks that UAV $j$ allocates to user $i$ with $v_{ij,k} \in \{1,0\}$. $\gamma_{ij,k}^{\text{u}} = \frac{P_U h_{ij}^k}{\sigma^2 + \sum_{l \in \mathcal{U}^k, l \neq j} P_U h_{il}^k}$ is the SINR between user $i$ and UAV $j$ over resource block $k$ with $\mathcal{U}^k$ being the set of users that use uplink resource blocks $k$ and $P_U$ is each user's transmit power.

Let $A$ be the data size of the tracking information. For user $i$, the total delay that consists of downlink and uplink transmission delays that can be given by:

$$D_{ij}(f_i, l_i, \boldsymbol{s}_{ij}, \boldsymbol{v}_{ij}, o_{ij}) = \frac{m(l_i, f_i)}{c_{ij}(\boldsymbol{s}_{ij}, e_{ij})} + \frac{A}{c_{ij}(\boldsymbol{v}_{ij})}, \quad (7)$$

where the first term is the time that UAV $j$ needs to transmit a VR image to user $i$ and the second term is the time that user $i$ needs to transmit the tracking information to UAV $j$.

### D. Problem Formulation

Given the defined system model, our goal is to develop an effective resource allocation scheme to allocate resource blocks, image quality level, and image format to maximize the QoE of all users while meeting the delay requirement of the VR system. However, this problem must jointly account for the coupled problems of user association and resource allocation. Moreover, the delay of each user depends on the decisions of all UAVs, due to interference and user association. Therefore, we formulate this problem as a noncooperative game [13] $\mathcal{G} = \left[\mathcal{B}, \{\mathcal{A}_j\}_{j \in \mathcal{B}}, \{\bar{u}_j\}_{j \in \mathcal{B}}\right]$. In this game, the players are the UAVs in set $\mathcal{B}$, $\mathcal{A}_j$ represents the action set of each UAV $j$, and $\bar{u}_j$ is the utility function of each UAV $j$. Here, each action $\boldsymbol{a}_j = [\boldsymbol{q}_j, \boldsymbol{v}_j, \boldsymbol{e}_j, \boldsymbol{l}_j, \boldsymbol{f}_j]$, consists of: (i) downlink licensed resource allocation vector $\boldsymbol{q}_j = \left[\boldsymbol{q}_{1j}, \boldsymbol{q}_{2j}, \ldots, \boldsymbol{q}_{U_jj}\right]$ with $U_j$ being the number of the users associated with UAV $j$, (ii) uplink resource allocation vector $\boldsymbol{v}_j = [\boldsymbol{v}_{1j}, \boldsymbol{v}_{2j}, \ldots, \boldsymbol{v}_{U_jj}]$, (iii) downlink unlicensed resource allocation vector $\boldsymbol{e}_j = \left[e_{1j}, e_{2j}, \ldots, e_{U_jj}\right]$, (iv) image quality allocation vector $\boldsymbol{l}_j = [l_{1j}, \ldots, l_{U_jj}]$, and (v) image format allocation vector $\boldsymbol{f}_j = [f_{1j}, \ldots, f_{U_jj}]$. Here, $l_{ij} \in \mathcal{L}, f_{ij} \in \mathcal{F}$, respectively, represent the image quality and format of user $i$. $e_{ij} \in \mathcal{M}$ where $\mathcal{M} = \left\{\vartheta, \frac{\vartheta}{2}, \ldots, \frac{\vartheta}{M}\right\}$ is a finite set of $M$-level fractions of UAV $j$'s total time that UAV $j$ can be used to transmit VR images. $\boldsymbol{q}_j$, $\boldsymbol{v}_j$ and $\boldsymbol{e}_j$ must satisfy:

$$\sum_{i \in \mathcal{U}_j} \boldsymbol{q}_{ij} = [1, 1, \ldots, 1], \quad \sum_{i \in \mathcal{U}_j} \boldsymbol{v}_{ij} = [1, 1, \ldots, 1], \quad (8)$$

$$\sum_{i \in \mathcal{U}_j} e_{ij} \leq 1, \quad e_{ij} \in \mathcal{M}, \quad (9)$$

where (8) indicates that each UAV will allocate all of its resource blocks to the associated users. We assume that each UAV $j$ chooses only one action at each time slot $t$. The utility function of each UAV $j$ can be given by:

$$u_j(\boldsymbol{a}_j, \boldsymbol{a}_{-j}) = \sum_{i \in \mathcal{U}_j} Q_i(l_{ij}, f_{ij}) \mathbb{1}_{\{D_{ij}(f_{ij}, l_{ij}, \boldsymbol{q}_{ij}, \boldsymbol{v}_{ij}, e_{ij}) \leq \gamma_D\}}, \quad (10)$$

where $\gamma_D$ represents the maximal tolerable delay for each VR user (maximum supported by the VR system being used), $\boldsymbol{a}_j \in \mathcal{A}_j$ represents an action of UAV $j$ and $\boldsymbol{a}_{-j}$ is the actions of all UAVs other than UAV $j$. In this game, the QoE and delay of all users

are analyzed over a period $T$ and, hence, the utility function is:

$$\bar{u}_j(\boldsymbol{a}_j, \boldsymbol{a}_{-j}) = \frac{1}{T} \sum_{t=1}^{T} u_j(\boldsymbol{a}_j, \boldsymbol{a}_{-j}). \quad (11)$$

Let $\pi_{j,\boldsymbol{a}_{ij}} = \frac{1}{T} \sum_{t=1}^{T} \mathbb{1}_{\{\boldsymbol{a}_{j,t} = \boldsymbol{a}_{ij}\}} = \Pr(\boldsymbol{a}_{j,t} = \boldsymbol{a}_{ij})$ be the probability of UAV $j$ using action $\boldsymbol{a}_{ij}$. Here, $\boldsymbol{a}_{j,t}$ is the action that UAV $j$ uses at time $t$ and $\boldsymbol{a}_{j,t} = \boldsymbol{a}_{ij}$ indicates that UAV $j$ adopts action $\boldsymbol{a}_{ij}$ at time $t$. $\boldsymbol{\pi}_j = \left[\pi_{j,\boldsymbol{a}_{1j}}, \ldots, \pi_{j,\boldsymbol{a}_{A_j j}}\right]$ is the action selection strategy of UAV $j$ with $A_j$ being the number of actions of UAV $j$. Based on the definition of the strategy, the utility function in (11) can be rewritten as follows:

$$\bar{u}_j(\boldsymbol{a}_j, \boldsymbol{a}_{-j}) = \sum_{\boldsymbol{a} \in \mathcal{A}} \left( u_j(\boldsymbol{a}_j, \boldsymbol{a}_{-j}) \prod_{k \in \mathcal{B}} \pi_{k,\boldsymbol{a}_k} \right), \quad (12)$$

where $\boldsymbol{a} \in \mathcal{A}$ with $\mathcal{A}$ being the action set of all UAVs. (12) actually captures the average utility value of each UAV $j$ over a period $T$.

Hence, for this model, our objective is to solve the proposed resource allocation game. One suitable solution is the mixed-strategy Nash equilibrium (NE), formally defined as follows [13]: A mixed strategy profile $\boldsymbol{\pi}^* = (\boldsymbol{\pi}_1^*, \ldots, \boldsymbol{\pi}_B^*) = (\boldsymbol{\pi}_j^*, \boldsymbol{\pi}_{-j}^*)$ is a *mixed-strategy Nash equilibrium* if, $\forall j \in \mathcal{B}$ and $\boldsymbol{\pi}_j$, we have:

$$\bar{u}_j(\boldsymbol{\pi}_j^*, \boldsymbol{\pi}_{-j}^*) \geq \bar{u}_j(\boldsymbol{\pi}_j, \boldsymbol{\pi}_{-j}^*), \quad (13)$$

where $\bar{u}_j(\boldsymbol{\pi}_j, \boldsymbol{\pi}_{-j}) = \sum_{\boldsymbol{a} \in \mathcal{A}} u_j(\boldsymbol{a}_j, \boldsymbol{a}_{-j}) \prod_{k \in \mathcal{B}} \pi_{k,\boldsymbol{a}_k}$ is the expected utility of UAV $j$ when it selects the mixed strategy $\boldsymbol{\pi}_j$. For our game, the mixed-strategy NE for the UAVs represents a solution of the game at which each UAV $j$ can maximize the QoE for its associated users while meeting the delay requirement of VR system, given the actions of its opponents.

## III. ECHO STATE NETWORKS FOR SELF-ORGANIZING RESOURCE ALLOCATION

Next, we introduce a reinforcement learning (RL) algorithm based on the neural network tools from echo state networks with leaky integrator neurons [9] that can be used to solve this game and find a mixed-strategy NE [13]. Traditional RL algorithms such as Q-learning [14] typically use a Q-table to record the utility values resulting from different actions and states. However, in our model, each action consists of five components which, if used in Q-learning, can result in an exponential increase in the number of actions and its corresponding utility values. Moreover, the update of Q-learning needs to traverse the Q-table to find the update position which will significantly decrease the update speed. Therefore, using a limited-size matrix to record all of the needed utility values as done in Q-learning may not be practical. In contrast, the proposed RL algorithm will use echo state networks as an approximation method to record the relationship between the actions, states and utility values and, hence, it can record all of the needed actions, states and utility values. Compared to the ESN algorithms in [5] and [6] that are used for discrete-time systems, the proposed ESN with leaky integrator neurons can dynamically adjust the update speed of the ESN's state and, hence, it can enable the UAVs to learn the continuous dynamics of their associated VR users.

TABLE I
ESN-BASED LEARNING ALGORITHM FOR RESOURCE ALLOCATION

**Inputs:** Mixed strategies $\boldsymbol{x}_{j,t}$
*Initialize:* $\boldsymbol{W}_j^{\text{in}}, \boldsymbol{W}_j, \boldsymbol{W}_j^{\text{out}}$, and $\boldsymbol{y}_j = 0$. Set $\boldsymbol{\pi}_{j,t}$ uniformly
   **for** each time $t$ **do**.
     (a) Estimate the value of the utility values $\hat{u}_{j,t}$ based on (16).
     **if** $rand(.) < \varepsilon$
       (b) Select action randomly.
     **else**
       (c) Choose action $\boldsymbol{a}_j = \arg\max_{\boldsymbol{a}_j \in \mathcal{A}_j} \hat{u}_{j,t}(\boldsymbol{a}_j)$.
     **end if**
     (d) Broadcast the index of the mixed strategy to other UAVs.
     (e) Receive the index of the mixed strategy as input $\boldsymbol{x}_{j,t}$.
     (f) Perform action to calculate the actual value of utility $\hat{u}_{j,t}$.
     (g) Set the mixed strategy $\boldsymbol{\pi}_{j,t}$ based on (14)
     (h) Update the dynamic reservoir state based on (15).
     (i) Update the output weight matrix based on (17).
   **end for**
**Output:** Prediction $\boldsymbol{y}_{j,t}$

### A. ESN Components

For each UAV, the proposed ESN algorithm consists of four components: (1) inputs, (2) ESN model, (3) actions, and (4) output, which are specified as follows:

• *Actions:* Each action of UAV $j$ is the vector $\boldsymbol{a}_j = [\boldsymbol{q}_j, \boldsymbol{v}_j, \boldsymbol{e}_j, \boldsymbol{l}_j, \boldsymbol{f}_j]$ that jointly considers the downlink licensed resource allocation, uplink licensed resource allocation, downlink unlicensed resource allocation, image quality allocation, and image format allocation.

• *Input:* The input of our ESN is defined as a vector $\boldsymbol{x}_{j,t} = [x_1, \cdots, x_B]^{\text{T}}$ where $x_j$ represents the index of the strategy of UAV $j$ at time $t$. To guarantee that any action always has a non-zero probability to be chosen, the $\varepsilon$-greedy exploration in [6] is used. This mechanism is also used to harmonize the tradeoff between exploitation and exploration. Therefore, the probability of UAV $j$ taking a certain action $\boldsymbol{a}_j$ can be given by:

$$\Pr(\boldsymbol{a}_j) = \begin{cases} 1 - \varepsilon + \frac{\varepsilon}{|\mathcal{A}_j|}, & \arg\max_{\boldsymbol{a}_j \in \mathcal{A}_j} \hat{u}_{j,t}(\boldsymbol{a}_j), \\ \frac{\varepsilon}{|\mathcal{A}_j|}, & \text{otherwise}, \end{cases} \quad (14)$$

where $\hat{u}_{j,t}(\boldsymbol{a}_j) = \sum_{\boldsymbol{a}_{-j} \in \mathcal{A}_{-j}} u_j(\boldsymbol{a}_j, \boldsymbol{a}_{-j}) \pi_{-j,\boldsymbol{a}_{-j}}$ represents the expected utility of a UAV $j$ with respect to the actions of other UAVs, $\pi_{-j,\boldsymbol{a}_{-j}} = \sum_{\boldsymbol{a}_j \in \mathcal{A}_j} \pi(\boldsymbol{a}_j, \boldsymbol{a}_{-j})$ represents the marginal probability distribution over the action set of UAV $j$, and $\mathcal{A}_{-j} = \prod_{k \neq j, k \in \mathcal{B}} \mathcal{A}_k$ denotes the set of actions other than UAV $j$.

• *Output:* The output of the ESN model at time $t$ is a vector of utility values $\boldsymbol{y}_{j,t} = [y_{j1,t}, y_{j2,t}, \ldots, y_{j|\mathcal{A}_j|,t}]$ where $y_{ji,t}$ is the estimated value of utility $\hat{u}_{j,t}(\boldsymbol{a}_{ji})$ due to action $i$ of UAV $j$. As time elapses, the ESN output, $\boldsymbol{y}_j$, will finally converge to the utility $\hat{\boldsymbol{u}}_j = [\hat{u}_j(\boldsymbol{a}_{j1}), \ldots, \hat{u}_j(\boldsymbol{a}_{j|\mathcal{A}_j|})]$.

• *ESN Model:* An ESN model is a learning architecture that can find the relationship between the input $\boldsymbol{x}_{j,t}$ and output $\boldsymbol{y}_{j,t}$, thus building the relationship between the UAVs' actions, strategies, and utility values. Mathematically, the ESN model consists of the output weight matrix $\boldsymbol{W}_j^{\text{out}} \in \mathbb{R}^{A_j \times (N_w + B + 1)}$, the input weight matrix $\boldsymbol{W}_j^{\text{in}} \in \mathbb{R}^{N_w \times B + 1}$, and the recurrent matrix $\boldsymbol{W}_j \in \mathbb{R}^{N_w \times N_w}$ with $N_w$ being the number of the units.

### B. ESN-Based Learning Algorithm for Resource Allocation

Next, we introduce the proposed ESN-based learning algorithm to find a mixed strategy NE. The ESN model can store historical ESN information such as input that can be used to find a fast converging process from the initial state to the mixed-strategy NE. The historical ESN information is stored in the reservoir state which can be given by:

$$\boldsymbol{\mu}_{j,t} = (1 - \delta C z) \boldsymbol{\mu}_{j,t-1} + \delta C f\left(\boldsymbol{W}_j \boldsymbol{\mu}_{j,t-1} + \boldsymbol{W}_j^{\text{in}} \boldsymbol{x}_{j,t}\right), \quad (15)$$

where $C$ is a time constant, $z$ is the leaking decay rate, $\delta$ is the step size, and $f(x) = \frac{e^x - e^{-x}}{e^x + e^{-x}}$ is the tanh function. In fact, (15) captures the difference between conventional ESN such as in [5] and [6] and the proposed ESN with leaky integrator neurons. When $\delta$ decreases, the ESN can be used for learning the continuous dynamics of VR users such as data rate and data size of a VR image. (15) shows that the reservoir state consists of the ESN input at time $t$ and the past dynamic reservoir state that includes the historical reservoir states and ESN inputs. Thus, the reservoir state can store the mixed strategy from time 0 to time $t$. Using the reservoir state with the output weight matrix, the proposed ESN algorithm can estimate the utility value $\hat{\boldsymbol{u}}_j$. The estimation of $\hat{\boldsymbol{u}}_j$ is:

$$\boldsymbol{y}_{j,t} = \boldsymbol{W}_{j,t}^{\text{out}} \left[\boldsymbol{\mu}_{j,t}; \boldsymbol{x}_{j,t}\right], \quad (16)$$

where $\boldsymbol{W}_{j,t}^{\text{out}}$ is the output weight matrix at time slot $t$. To enable the ESN to predict the utility value $\hat{\boldsymbol{u}}_j$ using reservoir state $\boldsymbol{\mu}_{j,t}$ and input $\boldsymbol{x}_{j,t}$, we must train the output matrix $\boldsymbol{W}_j^{\text{out}}$ using a linear gradient descent approach, which can be given by:

$$\boldsymbol{W}_{ji,t+1}^{\text{out}} = \boldsymbol{W}_{ji,t}^{\text{out}} + \lambda \left(\hat{u}_{ji,t} - y_{ji,t}\right) \boldsymbol{\mu}_{j,t}^{\text{T}}, \quad (17)$$

where $\boldsymbol{W}_{ji,t}^{\text{out}}$ is row $i$ of $\boldsymbol{W}_{j,t}^{\text{out}}$, $\lambda$ is the learning rate, and $\hat{u}_{ji,t}$ is the actual utility value. Here, $\hat{u}_{ji,t}$ is estimated by the utility value resulting from the actions performed by each UAV during each time slot $t$. Based on the above formulation, the distributed ESN-based learning algorithm performed by every UAV $j$ is summarized in Table I.

In this algorithm, each UAV can approximate the relationship between the actions, strategies, and utility values. During each iteration, the ESN-based algorithm can record the the strategies that each UAV uses and the corresponding utility values $\hat{\boldsymbol{u}}_j$. By scaling the input weight matrix $\boldsymbol{W}_j^{\text{in}}$ and input $\boldsymbol{x}_j$, the ESN-based algorithm can satisfy the convergence conditions of [6, Theorem 2]. As time elapses, each UAV $j$'s output resulting from action $i$ will converge to a final value $\hat{u}_{ji}$. At this convergence point, the output results $\hat{\boldsymbol{u}}_j$ indicates that the proposed algorithm converges to a mixed strategy-NE. This is due to the fact when $\hat{u}_j(\boldsymbol{a}_j)$ is maximum, $\Pr(\boldsymbol{a}_j)$ will be maximum ($\Pr(\boldsymbol{a}_j) = 1 - \varepsilon + \frac{\varepsilon}{|\mathcal{A}_j|}$). The strategy depends on $\hat{u}_j(\boldsymbol{a}_j)$. When the proposed learning algorithm finds the optimal $\hat{u}_j(\boldsymbol{a}_j)$, the strategy will be optimal and, hence, $\bar{u}_j(\boldsymbol{a}_j, \boldsymbol{a}_{-j})$ will be maximized.

## IV. SIMULATION RESULTS

In our simulations, a circular network area having a radius $r = 500$ m is considered with $U = 20$ uniformly distributed users and $B = 5$ uniformly distributed UAVs. The HTC VR device is considered in our simulations [15] and, hence, the rate requirement for VR transmission is set to 51.2 Mbits/s. The requirement of total delay that consists of the downlink and uplink transmission delay is 20 ms [16]. Other simulation parameters are listed in Table II. We compare our approach with: a) ESN algorithm in [5], b) Q-learning algorithm within an LTE-U, and c) Q-learning algorithm applied to an LTE system. All statistical results are averaged over 5000 independent runs.

TABLE II
SYSTEM PARAMETERS

| Parameter | Value | Parameter | Value | Parameter | Value |
|---|---|---|---|---|---|
| $P_B$ | 15 dBm | $V, S$ | 5, 5 | $M$ | 5 |
| $P_U$ | 20 dBm | $\eta_{\text{LoS}}^{\text{U}}, \eta_{\text{NLoS}}^{\text{U}}$ | 1.2, 23 | $z$ | 0.04 |
| $\sigma$ | -94 dBm | DIFS | 50 $\mu$s | $\varsigma$ | 20 dB |
| RTS | 352 $\mu$s | $E[S]$ | 1500 byte | $X$ | 11.9 |
| $F_u$ | 20 Mbit | ACK | 304 $\mu$s | $\delta$ | 1 |
| $C$ | 0.44 | SIFS | 16 $\mu$s | $K$ | 3 |
| CTS | 304 $\mu$s | $\xi$ | 4 Mbps | $Y$ | 0.13 |
| $F_C$ | 2 Gbit | $\eta_{\text{LoS}}^{\text{L}}, \eta_{\text{NLoS}}^{\text{L}}$ | 1, 20 | $N$ | 8 |

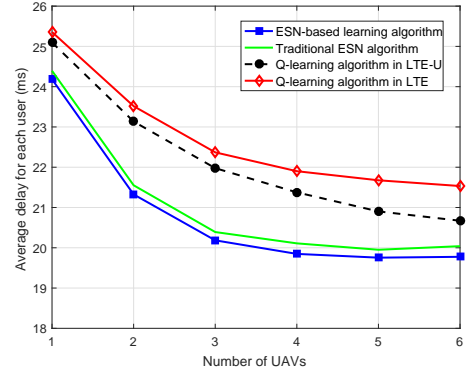

Fig. 3. Average delay of each user vs. number of UAVs.

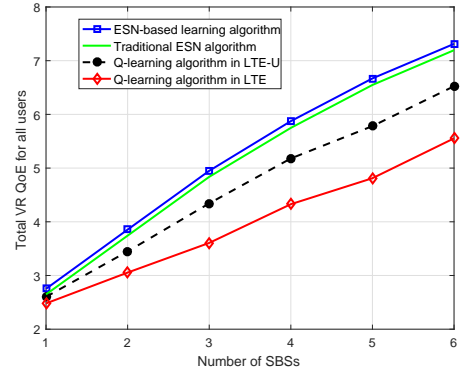

Fig. 4. Total VR QoE of all users vs. number of UAVs.

Fig. 3 shows how the average delay per user changes as the number of UAVs varies. From Fig. 3, we can see that as the number of UAVs increases, the average delay of all considered algorithms starts by first rapidly decreasing and then the decrease becomes slower. This is due to the fact that, as the number of UAVs increases, each user will have more choices of UAVs to connect to, and, hence, the distance between the UAVs and the users decreases. However, as the number of UAVs keeps increasing, the interference from the UAVs to the users increases. Fig. 3 also shows that when the delay reaches the delay requirement, the delay resulting from the proposed algorithm remains constant as the number of UAVs increases. This is due to the fact that our purpose is to maximize the QoE utility value while meeting the delay requirement of the VR system and, hence, once the delay requirement is met, the proposed scheme will no longer need to further reduce it.

In Fig. 4, we show how the total VR QoE for all users changes as the number of UAVs varies. From Fig. 4, we can see that, as the number of UAVs increases, the total QoE utility values of all considered algorithms will increase. This is due to the fact that, as the number of UAVs increases, the number of the users located in each UAV's coverage decreases and, hence, each UAV can allocate more resource to its associated users. Fig. 4 also shows that, for the case with 6 UAVs, the proposed algorithm can achieve, respectively, up to 14% and 27.1% gains in terms of total VR QoE for all users. These gains stem from the fact that the proposed algorithm can use the leaky integrator neurons to adjust the learning speed of ESNs. Hence, it enables the UAVs to learn the utility values resulting from the dynamical users and use the already learned utility value to estimate the new utility value that must be learned.

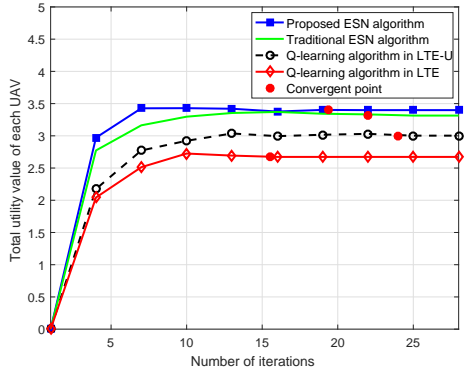

Fig. 5. Convergence of the proposed algorithm and Q-learning. Here, we use the ESN algorithm in [6] as a benchmark.

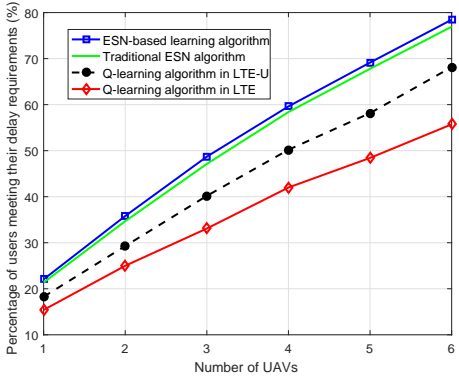

Fig. 6. Percentage of users meeting their delay requirements vs. number of UAVs.

Fig. 5 shows the number of iterations needed till convergence for all considered algorithms. In this figure, we can see that, as time elapses, the QoE utilities for both considered algorithms will increase until convergence to their final values. Fig. 5 also shows that the proposed algorithm yields up to 11.3% and 17.4% gains in terms of the number of the iterations needed to reach convergence compared to traditional ESN algorithm and Q-learning. This implies that the proposed algorithm can use the ESN with the leaky integrator neurons to adjust the learning speed and, hence, it can accurately approximate the relationship between the actions, strategies, and utility values.

Fig. 6 shows how the percentage of users meeting their delay requirements changes as the number of UAVs varies. From Fig. 6, we can see that the percentage of VR users meeting their delay requirements increases, as the number of UAVs increases. This stems from the fact that as the number of UAVs increases, the users have more UAV choices and the distances from the UAVs to the users decrease and, hence, the data rate of each user increases. Fig. 6 also shows that the proposed algorithm can achieve up to 40.1% gain in terms of the number of the users that meet the delay requirement. This is due to the fact the proposed algorithm use the ESN to record all the needed utility values to estimate the future utility values and allocate the unlicensed band to the users so as to increase the data rate of each user.

## V. CONCLUSION

In this paper, we have developed a novel framework that uses flying UAV-enabled networks to provide service for VR users in an LTE-U system. We have proposed a novel resource allocation framework for optimizing the users' QoE while meeting the delay requirement for each user. We have formulated this problem as a noncooperative game between the UAVs and we have developed a novel algorithm based on the machine learning tools of echo state networks with leaky integrator neurons. The proposed algorithm can dynamically adjust the update speed of the ESN's state and, hence, it can enable the UAVs to learn the continuous dynamics of their associated VR users. Simulation results have shown that the proposed approach yields significant gains in terms of total QoE utilities for all users compared to conventional approaches.